%% file: main.tex
\documentclass[conference,review,]{IEEEtran}
\IEEEoverridecommandlockouts

\usepackage{titlesec}
\usepackage[utf8]{inputenc}
\usepackage[T1]{fontenc}
\usepackage{newtxmath} 

\usepackage{amsmath,amssymb,amsfonts}
\usepackage{algpseudocode} 
\usepackage{amsmath}
\usepackage{array}
\usepackage{multirow}
\usepackage{graphicx}
\usepackage{caption}
\usepackage{subcaption}
\usepackage{balance}
\usepackage{algorithm}
\usepackage{algorithmicx}
\usepackage{esvect}
\usepackage{booktabs}
\usepackage[most]{tcolorbox}
\tcbuselibrary{theorems}
\newtcbtheorem{promptbox}{Prompt}
{colback=gray!10,boxrule=0pt,arc=2pt,
 left=5pt,right=5pt,top=3pt,bottom=3pt}{pr}

\usepackage{tabularx}
\usepackage{anyfontsize}
\usepackage[all]{nowidow}
\usepackage[normalem]{ulem}
\usepackage{xcolor}
\usepackage{colortbl}
\usepackage{enumitem}
\usepackage{makecell}
\usepackage[switch]{lineno}
\usepackage{framed}
\usepackage{wrapfig}

\usepackage{diagbox}

\usepackage{gensymb}

\newcolumntype{Y}{>{\centering\arraybackslash}X}

\definecolor{commentGray}{RGB}{120,120,120}
\renewcommand{\algorithmiccomment}[1]{\bgroup\color{commentGray}{//#1}\egroup}

\usepackage{pifont}
\usepackage{listings}
\usepackage{framed}
\definecolor{light-gray}{gray}{0.9}
\usepackage[framemethod=TikZ]{mdframed}
\mdfsetup{skipabove=5pt,skipbelow=3pt}
\usepackage{lipsum}
\mdfdefinestyle{RQFrame}{%
  linecolor=black,
  outerlinewidth=0.15pt,
  roundcorner=3pt,
  innertopmargin=2pt,
  innerbottommargin=2pt,
  innerrightmargin=4pt,
  innerleftmargin=4pt,
  backgroundcolor=light-gray}

\definecolor{javagreen}{rgb}{0.25,0.5,0.35} 

\lstdefinestyle{Alg}{
  basicstyle=\ttfamily\footnotesize,
  breaklines=true,
  tabsize=2,
  mathescape,
  numbers=left,
  xleftmargin=2.5em,
  xrightmargin=0.5em,
  frame=tb,
  framexleftmargin=2em,
  emph={Algorithm,Input,Output,for,each,do,if,else,Function,while,let,be,repeat,until,return,times,and,or,break,in,then,},
  emphstyle={\textbf},
  escapechar=?,
  morecomment=[l][\color{javagreen}]{//},
  columns=flexible,
}

\titleclass{\subsubsubsection}{straight}[\subsubsection]
\titleformat{\subsubsubsection}
  {\normalfont\normalsize\bfseries}{\thesubsubsubsection}{1em}{}
\titlespacing*{\subsubsubsection}
  {0pt}{3.25ex plus 1ex minus .2ex}{1.5ex plus .2ex}

\newcounter{subsubsubsection}[subsubsection]
\renewcommand\thesubsubsubsection{\thesubsubsection.\arabic{subsubsubsection}}

\newtheorem{definition}{Definition}

\newcommand\blfootnote[1]{%
  \begingroup
  \renewcommand\thefootnote{}\footnote{#1}%
  \addtocounter{footnote}{-1}%
  \endgroup
}

\begin{document}

\title{Towards Counterfactual Explanation and Assertion Inference for CPS Debugging
\thanks{
\IEEEauthorrefmark{1}All authors contributed equally to this work.}
}


\author{
\IEEEauthorblockN{1\textsuperscript{st} Zaid Ghazal\IEEEauthorrefmark{1}}
\IEEEauthorblockA{\textit{University of Michigan-Dearborn} \\
Dearborn, MI, USA \\
zghazal@umich.edu}
\and
\IEEEauthorblockN{2\textsuperscript{nd} Hadiza Yusuf\IEEEauthorrefmark{1}}
\IEEEauthorblockA{\textit{University of Michigan-Dearborn} \\
Dearborn, MI, USA \\
hyusuf@umich.edu}
\and
\IEEEauthorblockN{3\textsuperscript{rd} Khouloud Gaaloul\IEEEauthorrefmark{1}}
\IEEEauthorblockA{\textit{University of Michigan-Dearborn} \\
Dearborn, MI, USA \\
kgaaloul@umich.edu}

}
\maketitle

\blfootnote{\textcopyright~2026 IEEE.  Personal use of this material is permitted.  Permission from IEEE must be obtained for all other uses, in any current or future media, including reprinting/republishing this material for advertising or promotional purposes, creating new collective works, for resale or redistribution to servers or lists, or reuse of any copyrighted component of this work in other works.}

\begin{abstract}
Verification and validation of cyber-physical systems (CPS) via large-scale simulation often surface failures that are hard to interpret, especially when triggered by interactions between continuous and discrete behaviors at specific events or times. Existing debugging techniques can localize anomalies to specific model components, but they provide little insight into the input-signal values and timing conditions that trigger violations, or the minimal, precisely timed changes that could have prevented the failure. In this article, we introduce DeCaF, a counterfactual-guided explanation and assertion-based characterization framework for CPS debugging. Given a failing test input, DeCaF generates counterfactual changes to the input signals that transform the test from failing to passing. These changes are designed to be minimal, necessary, and sufficient to precisely restore correctness. Then, it infers assertions as logical predicates over inputs that generalize recovery conditions in an interpretable form engineers can reason about, without requiring access to internal model details. Our approach combines three counterfactual generators with two causal models, and infers success assertions. Across three CPS case studies, DeCaF achieves its best success rate with KD-Tree Nearest Neighbors combined with M5 model tree, while Genetic Algorithm combined with Random Forest provides the strongest balance between success and causal precision.
\end{abstract}






\section{Introduction}
\input{Sections/introduction}


\section{Problem Definition}
\label{sec:problem}
\input{Sections/problem}

\section{DeCaF Approach}
\label{sec:approach}
\input{Sections/approach}

\section{Evaluation}
\label{sec:evaluation}
\input{Sections/evaluation}

\section{Related Works}
\label{sec:related_work}
\input{Sections/related}

\section{Conclusion}
\label{sec:conclusion}
\input{Sections/conclusion}



\bibliographystyle{IEEEtran}
\bibliography{bibliography}

\end{document}

%% file: Sections/introduction.tex
Recent years have seen significant growth in simulation-based testing of cyber-physical systems (CPS)~\cite{10.1145/3711906,khatiri2023simulation,arrieta2019pareto}, driven by the need to validate increasingly complex systems with both discrete and continuous dynamics. Beyond academic research, industry and government have embraced simulation at scale. For example, the U.S. Air Force’s Model One initiative~\cite{wsj_digitaltwins2024} integrates 50 military simulations into a unified digital-twin framework as a strategic tool for testing. Simulation-based testing of CPS often relies on industrial-strength modeling environments like MATLAB/Simulink~\cite{chaturvedi2017modeling}, which enable engineers to test system designs in stochastic and dynamic environments. In such conditions, testing frequently surfaces a wide variety of failures that are difficult to interpret or explain due to continuous-time behaviors and intricate interactions among multiple inputs. Failures may occur only under precise signal values, timing conditions, or rare sequences of state transitions~\cite{wehbe2025finding}. Several automated methods have been proposed to assist with failure diagnostics~\cite{sampath2002failure}, including trace monitoring and diagnostics~\cite{ferrere2015trace,bartocci2018localizing,liu2016simulink}, model-based structural analysis~\cite{bartocci2021cpsdebug,deng2023causal}, and statistical or causal debugging~\cite{podgurski2020counterfault,johnson2020causal,chadbourne2023applications,chakraborty2019root,ren2019root}. These methods can flag anomalous components or trace fault propagation, but they offer little insight into the input conditions that trigger failures. 

Existing debugging approaches predominantly operate at the level of model structure or internal signals. For example, Bartocci et al.~\cite{bartocci2021cpsdebug,bartocci2018localizing} provide explanations that are rooted in model-level properties, by analyzing failing traces and identifying root events and their propagation across model elements, to locate anomalous Simulink components or execution paths responsible for violations. Deng et al.~\cite{deng2023causal} introduce a causal temporal logic that explains fault propagation in Simulink/Simscape models. However, none of these works support input-level explanations of failures. Recent works~\cite{chadbourne2023applications,johnson2020causal,podgurski2020counterfault,chakraborty2019root,ren2019root} have explored causal reasoning as a means for fault localization in complex systems and programs. Johnson et al.~\cite{johnson2020causal} propose a causal testing method to identify root software defects in Java programs, using counterfactual reasoning to explain failures by showing what changes in the input would prevent them. Yet, this work mainly targets discrete unit-level programs. To the best of our knowledge, no existing approach has applied causal reasoning to CPS with dynamic behaviors, where failures arise only under specific input conditions in terms of signal values and their timing, that capture complex continuous–discrete interactions. The closest effort, CaSTL~\cite{CaSTL}, integrates causal theory with signal temporal logic to generate fault explanations. Debugging CPS requires separating true system bugs from test formulation issues, such as violated environment assumptions or an exceeded operational design domain. Whereas methods like CaSTL diagnose component-level faults in a white-box setting, DeCaF acts as a first-line tool for offline fault diagnosis by enabling black-box recovery at the input level.

Further, recent studies on assertion-based inference~\cite{kapugama2022human,gopinath2020abstracting,soremekun2020inputs,kampmann2020does,gaaloul2021combining,enrich} have focused on generating assertions that characterize the circumstances under which the system-level behavior satisfies functional requirements. For example, Daikon~\cite{ernst2007daikon} uses dynamic invariant detection to report properties observed across program executions, while Alhazen~\cite{kapugama2022human,soremekun2020inputs} relies on grammar mining to predict failures for given inputs and to produce additional failure-inducing inputs. These approaches can support program understanding and input prediction, but lack the capacity for causal reasoning, i.e., answering `what-if' questions such as: \textit{What specific change in input signals would have prevented this failure? Within which time intervals would altered inputs restore correctness?} Gaaloul et al.~\cite{gaaloul2021combining,miningassumption} similarly focus on synthesizing environment assumptions in Simulink models but do not provide input-level counterfactual explanations of CPS failures.

\begin{figure*}
\centering
\includegraphics[width=.95\textwidth]{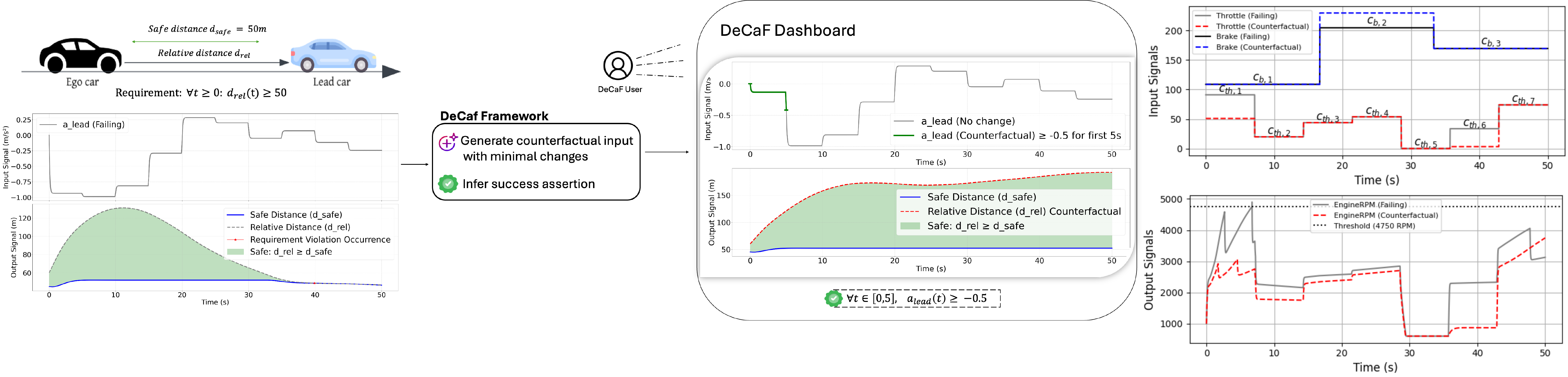}
\caption{Example Illustration of DeCaF and Example signals of counterfactual explanation generated for the AT case-study systems with Throttle and Brake inputs.}
\label{fig:example}
\end{figure*}

In this paper, we introduce DeCaF, a framework that supports debugging of CPS modeled in Simulink by (i) generating counterfactual-guided explanations of failure-inducing inputs, and (ii) inferring assertions that characterize the conditions under which correctness is restored. When a CPS model fails during simulation, debuggers can often trace the problem to internal components but provide limited insight into which input values or timing conditions triggered the violation, or how these inputs (or their characterization as additional preconditions to the test) could be adjusted to restore correctness. 
Figure~\ref{fig:example} shows DeCaF on an adaptive cruise control Simulink model~\cite{pananurak2009adaptive} with a safe distance requirement, $d_{rel}\ge d_{safe}=50,m$. When a simulation violates the requirement, DeCaF generates a counterfactual input signal that avoids the failure and infers an interpretable success assertion over the input, for example $\forall t\in[0,5],, a_{lead}(t)\ge -0.5$, helping engineers understand recovery conditions without internal access or extra instrumentation. DeCaF is designed for scenarios where immediate system-level interventions are impractical, due to operational continuity, cost, or lack of internal system-model access.

\textbf{Contributions.} Our contributions are as follows:
\begin{itemize}
    \item[(i)] We introduce DeCaF, a novel counterfactual-guided explanation and characterization framework for debugging CPS modeled in Simulink. Given a failing test case, DeCaF generates interpretable counterfactuals as minimal changes to input signal values or timing intervals that are both necessary and sufficient to restore correctness.
    
    \item[(ii)] We infer assertion-based characterizations from the generated counterfactuals that generalize the input conditions, guaranteeing correct behavior. These assertions can guide requirement refinement in uncontrolled environments and enable recovery reasoning without access to internal system components.
    
    \item[(iii)] We formalize the problems of counterfactual generation and assertion inference for CPS, and introduce a transformation mechanism that maps learned explanations, expressed in terms of control points, into concrete input-signal modifications and formal temporal assertions.
    
    \item[(iv)] We conduct an empirical evaluation of DeCaF on three CPS case studies covering multiple requirements. We consider six configurations of DeCaF, by combining three counterfactual generation strategies and two ML models. The evaluation measures the success, necessity, and sufficiency of DeCaF’s counterfactuals in correcting faulty behaviors, as well as the complexity and generalizability of the inferred assertions.
\end{itemize}

\textbf{Organization.} Section~\ref{sec:problem} defines the problem of counterfactual-guided explanation and assertion-based characterization for CPS, illustrated by examples. Section~\ref{sec:approach} introduces the DeCaF framework and details its four stages. Section~\ref{sec:evaluation} presents the metrics and experimental evaluation of DeCaF. Section~\ref{sec:related_work} compares our work with related research. Section~\ref{sec:conclusion} concludes the article.

%% file: Sections/problem.tex
In this section, we formally define the problem of counterfactual-guided explanation and assertion-based characterization addressed by DeCaF, illustrating the concepts with examples. We then present the transformation rules that map the generated counterfactuals and inferred success assertions, expressed in terms of control points, into concrete input signals and their corresponding assertions.

Figure~\ref{fig:example} illustrates the automatic transmission (AT) system, a representative CPS model that controls gear shifts based on vehicle speed and throttle input. $\mathit{Throttle(Failing)}$ and $\mathit{Brake(Failing)}$ denote the failing test inputs, while $\mathit{Throttle(Counterfactual)}$ and $\mathit{Brake(Counterfactual)}$ denote the corresponding counterfactual inputs generated by DeCaF. The system output $\mathit{EngineRPM}$ refers to the engine speed. Under the original inputs, the output $\mathit{EngineRPM (Failing)}$ violates the 4750 RPM threshold, whereas the counterfactual inputs restore correctness of the system output $\mathit{EngineRPM (Counterfactual)}$ by satisfying the threshold.  A common CPS testing approach encodes input signals using control points~\cite{tuncali2019requirements, arrieta2019pareto}, where signals are represented by sequences of values at fixed time intervals, interconnected using a piecewise constant interpolation. In this example, $\mathit{Throttle}$ is represented with 7 control points, each assigned a value in $[0,100]$, and $\mathit{Brake}$ with 3 control points in $[0,325]$. As shown, the input $\mathit{Throttle(Failing)}$ and $\mathit{Brake(Failing)}$ trigger a failure since $\mathit{engineRPM}$ exceeds 4750 within 10 seconds. DeCaF generates a counterfactual test input (i.e., $\mathit{Throttle(Counterfactual)}$ and $\mathit{Brake(Counterfactual)}$) by modifying the first and sixth $\mathit{Throttle}$ points and the second $\mathit{Brake}$ point. It further infers an assertion: $(\forall t \in [0,7.14],\ \mathit{Throttle}(t) \leq 54.50) \ \land\ (\forall t \in [35.71,42.46],\ \mathit{Throttle}(t) \leq 59.30) \ \land\ (\forall t \in [16.67,33.33],\ \mathit{Brake}(t) \geq 212.30$), which characterizes recovery conditions that turn a failing test input into a passing one.


\begin{definition}[\textbf{Control Point-Based Signal}]
\label{def:cp}
Let $u: \mathbb{T} \rightarrow \mathbb{R}$ be an input signal. 
We encode $u$ using $n_u$ control points, i.e., $c_{u,0}, c_{u,1}, \ldots, c_{u,n_u-1}$, 
equally distributed over the time domain $\mathbb{T}=[0,b]$, positioned at a fixed time distance $I=\frac{b}{n_u-1}$.
Each control point $c_{u,y}$ contains the value of the signal $u$ at time instant $y \cdot I$, 
for $y = 0, 1, \ldots, n_u-1$. 
\end{definition}
\noindent \textbf{Example.} In Figure~\ref{fig:example}, the $\mathit{Throttle}$ signal (denoted $th$) and $\mathit{Brake}$ signal (denoted $b$) are represented using $n_{th}=7$ and $n_b=3$ control points, respectively (i.e., $(c_{th,1}, \ldots, c_{th,7})$ and $(c_{b,1}, c_{b,2}, c_{b,3})$), where each control point determines the value of the signal in the time domain $[0,50s]$.\\

The Simulink model $M$ of a CPS is encoded using a control-point representation of the form 
$\langle \text{int}_u, R_u, n_u \rangle$, where 
$\text{int}_u$ is an interpolation function (e.g., linear, piecewise constant, cubic), 
$R_u \subseteq \mathbb{R}$ is the domain of admissible values, 
and $n_u$ is the number of control points for signal $u$. 
To generate the signal $u(t)$, we place $n_u$ control points at uniformly spaced intervals 
and assign values $c_{u,1}, \ldots, c_{u,n_u} \in R_u$, 
interpolating them with $\text{int}_u$.

\begin{definition}[\textbf{Test Input and Counterfactual}]
\label{def:counterfactual}
Let $M$ be a Simulink model, and let $\varphi$ be a requirement over its behavior. 
Let $\mathcal{U}$ denote the set of all possible test inputs for $M$, 
where each test input is a tuple of $m$ signals $\mathbf{u} = (u_1, u_2, \ldots, u_m), \quad u_i: \mathbb{T} \to \mathbb{R}$, with each $u_i$ encoded as in Definition~\ref{def:cp}.  
Assume that $\mathbf{u}$ is a failure-inducing input such that $M(\mathbf{u}) \not\models \varphi$. 
A \emph{counterfactual} $\mathbf{u}^*$ is a modified test input obtained from $\mathbf{u}$ 
that restores satisfaction, i.e., $M(\mathbf{u}^*) \models \varphi$. 
\end{definition}
\noindent \textbf{Example.} In Figure~\ref{fig:example}, 
the signals $\mathit{Throttle(Counterfactual)}$ and $\mathit{Brake(Counterfactual)}$ 
jointly form a counterfactual test input $\mathbf{u}^*$ corresponding to the failing input $\mathbf{u}$ 
with $\mathit{Throttle(Failing)}$ and $\mathit{Brake(Failing)}$ signals. 

\begin{definition}[\textbf{Assertion}]
\label{def:assertion}
An \textit{assertion} for a system model $M$ is a logical expression of the form $\mathit{condition} \Rightarrow \mathit{verdict}$, where $\mathit{condition}$ is a predicate over the inputs of $M$, and $\mathit{verdict}$ denotes the expected execution outcome of $M$ on any input $u$ that satisfies $\mathit{condition}$. An assertion of the form $\mathit{condition} \Rightarrow \mathit{fail}$ indicates that $M$ configured with $u$ is expected to fail $\varphi$ if it satisfies $\mathit{condition}$. Likewise, $\mathit{condition} \Rightarrow \mathit{pass}$ indicates that $M$ configured with $u$ is expected to pass $\varphi$ if it satisfies $\mathit{condition}$. We refer to assertions with a $\mathit{fail}$ verdict as \emph{fail assertions}, and those with a $\mathit{pass}$ verdict as \emph{success assertions}. A \emph{success assertion} for $\mathcal{CF}$ is an assertion satisfied by every counterfactual input $\mathbf{u}^* \in \mathcal{CF}$ defined by Definition~\ref{def:counterfactual}.
\end{definition}

\noindent Let $A$ be an assertion of the form $\mathit{condition} \Rightarrow \mathit{verdict}$. $\mathit{condition}$ is defined over the control-point inputs and can be represented in a canonical form: $condition ::= C_1 \lor C_2 \lor \dots \lor C_n, \quad C_i \in \mathcal{C}$, where each $C_i$ is a conjunction of predicates over control points: $C_i ::= P_1(c_{j1}) \land P_2(c_{j2}) \land \dots \land P_m(c_{jm})$. Each predicate $P_k(c_{jk})$ constrains the value assigned to a control point $c_{jk}$ for an input signal $u_j$. 

\noindent \textbf{Example.} A success assertion characterizing the recovery conditions for the failing test input in Figure~\ref{fig:example} is: $(c_{th,1} \leq 54.50) \land (c_{th,6} \leq 59.30) \land (c_{b,2} \geq 212.30) \Rightarrow pass$, where the predicates are satisfied by the counterfactual signals $\mathit{Throttle(Counterfactual)}$ and $\mathit{Brake(Counterfactual)}$ generated by DeCaF.

We formulate the counterfactual-guided explanation problem as follows. Let $\mathcal{U} = \{(\mathbf{u}_1, y_1), \ldots, (\mathbf{u}_n, y_n)\}$ be a labeled test suite. Each $\mathbf{u}_i$, where $i \in [1, n]$, is a control-point–based test input and $y_i \in \{\textsf{pass}, \textsf{fail}\}$ is its verdict with respect to requirement $\varphi$. For each failing input $\mathbf{u}_i$ such that $y_i = \textsf{fail}$, find:
\begin{itemize}
\item A set of counterfactuals $\mathcal{CF}_i = \{\mathbf{u}_{i1}^*, \ldots, \mathbf{u}_{im}^*\}$. 
\item A success assertion $A_i$ such that $A_i(\mathbf{u}) \Rightarrow \textsf{pass}$ for all $\mathbf{u}_{ij}^* \in \mathcal{CF}_i$.
\end{itemize}

Note that DeCaF operates on control point-based test inputs defined in Definition~\ref{def:cp}. To enable interpretability of the counterfactual-guided explanation and assertion-based characterization, these control points are \emph{concretized} into the actual signals over the system inputs by applying an interpolation function over the sequence of control points to form a signal~\cite{tuncali2019requirements, arrieta2019pareto,  miningassumption} and execute the system under test. For example, in Figure~\ref{fig:example}, the signal $\mathit{Throttle(Failing)}$ is constructed as a piecewise constant function over the simulation domain, with each constant segment defined by one of the 7 control points. We assume that system inputs follow piecewise constant interpolation. This is consistent with industrial CPS benchmarks~\cite{cruisecontroller,clutchlockup,guidancecontrol,dcmotor} and with the ARCH benchmark~\cite{khandait2024arch}, which compares piecewise continuous and constant functions. Results from ARCH-COMP19~\cite{ernst2019arch} show minimal performance differences between the two, indicating that our assumption is not restrictive.


Success assertions are generated as conditions that are conjunctions of relational expressions over control points. To translate the assertions derived over control points into concrete assertions over system variables, we follow one translation rule that operates in two steps. First, for each relational expression over a control point $c_{u,j}$, we introduce a universal quantifier over the corresponding time interval $[j\cdot I, (j+1)\cdot I]$, replacing $c_{u,j}$ with the signal value $u(t)$. Second, we apply logical translation to combine universal quantifiers across conjunctive expressions. For example, the counterfactual in Figure~\ref{fig:example} satisfies an assertion in the control-point-level form:
$(c_{th,1} \leq 54.50) \land (c_{th,6} \leq 59.30) \land (c_{b,2} \geq 212.30)$, 
which is translated into the signal-level form: $(\forall t \in [0,7.14]: \mathit{Throttle}(t) \leq 54.50) \wedge\ (\forall t \in [35.71,42.46]: \mathit{Throttle}(t) \leq 59.30) \wedge\ (\forall t \in [16.67,33.33]: \mathit{Brake}(t) \geq 212.30)$.

%% file: Sections/approach.tex
\begin{figure}[t]
    \centering
    \includegraphics[width=1\linewidth]{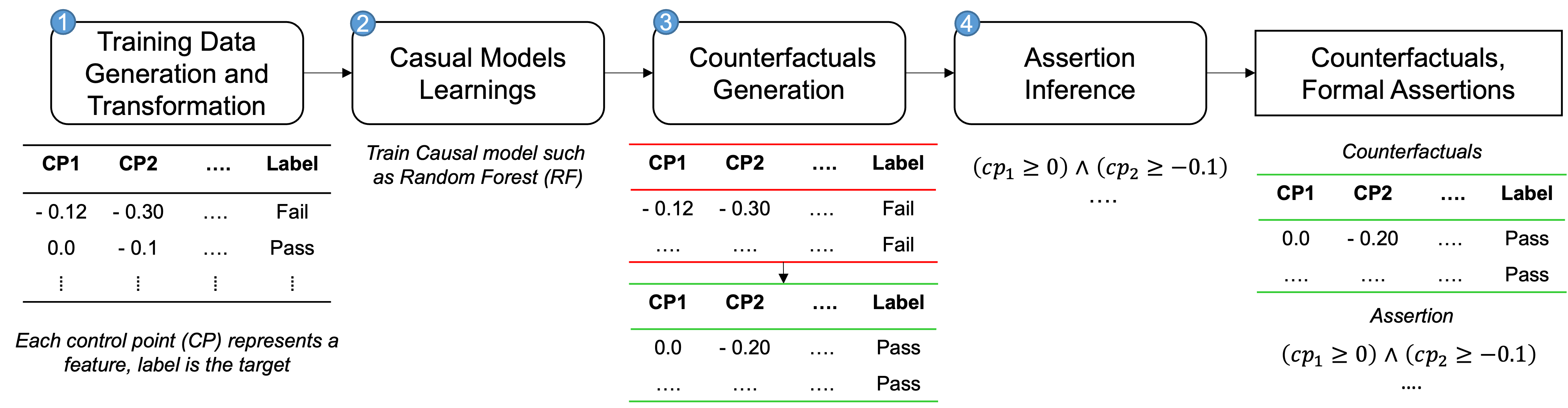}
    \caption{Overview of DeCaF Framework.}
     \label{fig:decaf_overview}
\end{figure}

In this section, we present DeCaF, a counterfactual-guided explanation and assertion-based characterization framework designed to support debugging of CPS. As a prerequisite, the system under test (SUT) is accompanied by a specification document that defines the input variables, their data types, and valid operating ranges. The approach takes as input a Simulink-based model of the SUT, a formalized system requirement, and a set of system-level input variables with value ranges extracted from the specification. Figure~\ref{fig:decaf_overview} shows an overview of DeCaF's framework. DeCaF first
generates training data, then trains causal models, generates counterfactuals, and finally infers
success assertions as defined in Definition~\ref{def:assertion}. The output is a report summarizing the failing test inputs, their counterfactuals, and the corresponding success assertions. The framework follows four steps.

\subsection*{Step~1: Training data generation and transformation}
\label{sec:step1}

\begin{algorithm}
\caption{Simulated Annealing for Data Generation (Minimization)}
\label{alg:sa}
\renewcommand{\algorithmicrequire}{\textbf{Input:}}
\renewcommand{\algorithmicensure}{\textbf{Output:}}
\begin{algorithmic}[1]\small
\Require $M$: SUT model, $\varphi$: STL specification, $S$: Input ranges
\Require $t>0$: Initial temperature, $\alpha\in(0,1)$: Cooling rate, $N$: Max iterations
\Ensure $u_{best}$ (best input), $v_{best}$ (its verdict)

\State $u \gets \textsc{Random}(S)$; $rb \gets \rho_\varphi(M(u),\varphi)$ 
    \Comment{Initialization}
\State $u_{best} \gets u$; $rb_{best} \gets rb$ 
    \Comment{best input track}
\For{$i=1$ to $N$}
  \State $u' \gets \textsc{Tweak}(u)$; $rb' \gets \rho_\varphi(M(u'),\varphi)$ 
    \Comment{Generate input}
  \If{$rb' < rb$ \textbf{or} $\textsc{Rand}(0,1) < e^{-(rb'-rb)/t}$}
    \State $u \gets u'$; $rb \gets rb'$ 
      \Comment{Probability-based selection}
  \EndIf
  \If{$rb < rb_{best}$} 
    $u_{best} \gets u$; $rb_{best} \gets rb$ 
      \Comment{Update best}
  \EndIf
  \State $t \gets \alpha \cdot t$ 
    \Comment{Decrease temperature according to cooling rate}
\EndFor
\State $v_{best} \gets \texttt{fail}$ if $rb_{best} < 0$, else \texttt{pass}$ $ 
    \Comment{Assign verdict}
\State \Return $u_{best}, v_{best}$ 
    \Comment{Return best input and its verdict}
\end{algorithmic}
\end{algorithm}

This step uses Simulated Annealing~\cite{metaheuristicsbook} to generate training inputs for the SUT, as shown in Algorithm~\ref{alg:sa}. 
Simulated Annealing is a metaheuristic search algorithm that iteratively perturbs candidate inputs and evaluates them against a formal requirement $\varphi$, specified in Signal Temporal Logic (STL), a language for expressing temporal properties of real-valued signals in cyber-physical systems~\cite{maler2004monitoring}. It occasionally accepts worse candidates with a probability that decreases as the temperature $t$ cools, to balance exploration with exploitation. 
Inputs are represented as control-point encodings, transformed into actual signals over the system inputs, and simulated on the SUT. The algorithm returns $(u_{best}, v_{best})$, where $u_{best}$ is the input with the minimal robustness $rb$ found and $v_{best}\in\{\texttt{pass},\texttt{fail}\}$ is its verdict determined by robustness: \texttt{fail}\ if $rb<0$ (the requirement $\varphi$ is violated), and pass otherwise. By executing the algorithm repeatedly, we obtain a labeled set of inputs data that forms the supervised training dataset ($TS$) for causal models.

\subsection*{Step~2: Causal model learning}
This step trains a causal model ($CM$, Line~\ref{alg:cf:line2}) that approximates the behavior of the SUT by learning from the labeled test inputs generated in Step 1. The algorithm starts by transforming the TS generated in Step~1 to reformat each input trace in the dataset into a representation where each input control point forms an input feature for the causal model, and the verdict is the outcome label (\texttt{pass} or \texttt{fail}). The model captures correlations between the test input data and their labels. We use the learned model only to guide search and ranking, while every candidate counterfactual is ultimately validated by replaying it in the simulator and checking the requirement robustness, which remains the source of truth. The causal model is trained using the following machine learning (ML) techniques. 
\begin{itemize}
    \item \emph{M5 model tree}~\cite{Quinlan1992, Holmes1999} recursively partitions the input space to form a decision tree where each path corresponds to a set of input conditions, and each leaf holds a regression function. M5 is particularly characterized by its ability to produce human-readable decision rules where each path in the tree is an interpretable explanation of the assigned verdicts.
    \item \emph{Random Forest}~\cite{biau2016random} builds an ensemble of decision trees trained on bootstrapped subsets of the data. Each tree learns a different perspective of the input-output relationships, and the ensemble aggregates their predictions to increase robustness. 
    \item \emph{Support Vector Machine (SVM)}~\cite{CERVANTES2020189} constructs a hyperplane or set of hyperplanes in a high-dimensional space to optimally separate passing from failing test inputs. SVM attempts to maximize the margin between verdict classes.
    \item \emph{RIPPER}~\cite{cohen1995fast} incrementally builds a set of propositional rules to classify test input data. It constructs rules greedily to cover positive examples while avoiding negatives, prunes them for generality, and iteratively refines the rule set to improve accuracy. The output of RIPPER is a compact collection of human-interpretable if-then rules.
\end{itemize}
The technique is set in configuration using a model type specified in configuration $C.modelType$ 
(Line~\ref{alg:cf:line2}). The resulting causal model is then passed to the subsequent step. The algorithm then identifies failing inputs by querying the causal model or labels to extract test inputs violating requirements (Line~\ref{alg:cf:line3}). These become targets for counterfactual generation to discover minimal modifications that yield passing outcomes.

\begin{algorithm}
\caption{Counterfactual-Guided Explanation and Assertion-Based Inference}
\label{alg:counterfactual}
\renewcommand{\algorithmicrequire}{\textbf{Input:}}
\renewcommand{\algorithmicensure}{\textbf{Output:}}
\begin{algorithmic}[1]\small
\Require $\mathit{TS}$: Training set of labeled control-point inputs, $C$: Configuration parameters
\Ensure $CF$: Selected set of best counterfactuals, $A$: Inferred success assertion

\State $\mathit{D} \gets \textsc{Transform}(TS)$ \label{alg:cf:line1}
\State $\mathit{CM} \gets \textsc{LearnCausalModel}(D, C.modelType)$ \label{alg:cf:line2}
\State $Failures \gets \textsc{IdentifyFailingInputs}(D, \mathit{CM})$ \label{alg:cf:line3}
\State $CF_{t} \gets \emptyset$ \label{alg:cf:line4}
\For{each $x_f \in Failures$} \label{alg:cf:line5}
    \State $CF_{x_f} \gets \textsc{GenerateCounterfactuals}(x_f, \mathit{CM}, C)$ \label{alg:cf:line6}
    \State $CF_{x_f} \gets \textsc{EvaluateCF}(CF_{x_f}, objectives)$ \label{alg:cf:line7}
    \State $CF_{t} \gets CF_{t} \cup CF_{x_f}$ \label{alg:cf:line8}
\EndFor \label{alg:cf:line9}
\State $CF \gets \textsc{SelectCounterfactuals}(CF_{t}, C.selectionCriterion)$ \label{alg:cf:line10}
\State $A \gets \textsc{InferSuccessAssertion}(CF, \mathit{CM})$ \label{alg:cf:line11}
\State \textsc{DeployAssertion}($A$) \label{alg:cf:line12}
\State \Return $CF, A$ \label{alg:cf:line13}
\end{algorithmic}
\end{algorithm}

\subsection*{Step~3: Counterfactual generation}
This step takes as input the trained causal model ($CM$), the set of failing inputs ($Failures$), and it generates a set of optimal counterfactuals (see Definition~\ref{def:counterfactual}) that (i) change the verdict from \texttt{fail} to \texttt{pass}, (ii) remain close to the original input (proximity), and (iii) are diverse and plausible. To do so, the framework initializes an empty set to store generated counterfactuals ($CF_t$) and iterates through each failing input (Line~\ref{alg:cf:line5} to Line~\ref{alg:cf:line9}) to generate candidate counterfactuals using one of the three configured strategies: Genetic Algorithm (GA), Random Search (RS), and KD-Tree Nearest Neighbors (KD). 
Once all failing test inputs are parsed, the algorithm selects the best counterfactuals per failing input (Line~\ref{alg:cf:line10}). This step returns the set of best counterfactuals ($CF$). Our approach incorporates the following model-agnostic counterfactual generation (execution in Line \ref{alg:cf:line6}) strategies:
\begin{itemize}
    \item \emph{Random Search (RS)} explores the input space by applying random perturbations to the control points of failing inputs. RS is fully stochastic and allows constraints on which features can be modified, as well as bounds on input ranges.
    \item A \emph{Custom Genetic Algorithm (GA)}, known as GeCo~\cite{GeCO}, is used to evolve candidate counterfactuals by uniquely partitioning its population into feasible and infeasible groups. For the feasible group, candidates that successfully flip the model’s prediction to pass, the algorithm’s primary goal is to minimize the distance from the original input. In parallel, for the infeasible group, the evolutionary operators focus on pushing candidates across the decision boundary to make them valid. This dual-objective strategy allows the algorithm to efficiently balance the search between finding a valid counterfactual and optimizing its cost.
    \item \emph{KD-Tree Nearest Neighbors (KD)} leverages the principles of prototype-guided explanations. It constructs a KD-Tree from the training dataset to perform an extremely fast nearest neighbor search. For a given failing input, this method identifies the closest data points that belong to the desired passing class. These neighbors serve as data-driven prototypes and are returned as the top-k candidate counterfactuals~\cite{van2021interpretable}. 
\end{itemize}

\subsection*{Step~4: Success assertion inference}
\label{subsection stage 4}
This step takes as input the selected set of best counterfactuals ($CF$) generated in Step~3 and infers a success assertion ($A$) (Line~\ref{alg:cf:line11}) that characterizes the input changes that allowed the set of counterfactuals to steer the system from violating to satisfying the requirement. Success assertions are not meant to define passing conditions over the entire input space. Instead, they generalize the behavioral pattern of generated counterfactuals, capturing in a set of interpretable assertions, the input conditions that distinguish them from the failing input. In essence, the assertion provides a concise, human-readable conditions on input variables and time under which the success behavior of the SUT is recovered. To infer these assertions, DeCaF applies the supervised learning algorithm \mbox{M5}~\cite{Quinlan1992}, well-suited for high-dimensional continuous input signals of cyber-physical systems. M5 builds a model tree by recursively partitioning the input space to minimize within-node variance, where each root-to-leaf path corresponds to a conjunction of relational expressions over input variables. These paths yield interpretable rules describing how the failing test can be repaired and may later be deployed for runtime monitoring or corrective actions. After constructing the tree, redundant predicates and overlapping intervals are pruned, and identical consequents are merged to produce the assertion as specified in Definition~\ref{def:assertion}. The inferred assertion is then optionally deployed within the SUT for runtime monitoring or automatic repair (Line~\ref{alg:cf:line12}). Finally, the algorithm returns the selected set of best counterfactuals ($CF$) together with the success assertion ($A$).

%% file: Sections/evaluation.tex
We evaluate DeCaF by answering the following research questions:
\begin{itemize}
    \item \textbf{RQ1 -- Effectiveness:} How effective is DeCaF at generating valid counterfactuals for debugging CPS failures? We evaluate the success of DeCaF in generating valid counterfactuals that resolve system failures across various generation strategies and ML models.
    \item \textbf{RQ2 -- Goodness:} How necessary and sufficient are DECaF’s generated counterfactuals for fixing the faulty system behavior? We evaluate the goodness of DeCaF’s counterfactuals by assessing their necessity and sufficiency in restoring the correct system behavior.
    \item \textbf{RQ3 -- Complexity vs. Generalizability:} What is the trade-off between complexity and generalizability in DeCaF’s inferred assertions? We analyze how concise (predicate complexity) and generalizable (input space coverage) the inferred success assertions are.
\end{itemize}

\subsection{Study Subjects and Experimental Settings}
\label{sec:experimental_settings}

We use Simulink models of three case-study systems detailed as follows: 

\begin{itemize}
    \item \textit{Automatic Transmission (AT).} The system is detailed as a motivating example in Section~\ref{sec:problem}, consisting of a controller that selects gears from 1 to 4 based on throttle and brake inputs, engine load, engine speed, and vehicle speed. The model was developed by MathWorks~\cite{hoxha2014benchmarks} and has become a widely used benchmark in CPS verification research.
    \item \textit{Adaptive Cruise Control (ACC).} The model of the adaptive cruise control system is published by Mathworks~\cite{mathworks2021acc}. This system simulates an ego car and a lead car operating in a controlled environment, where the goal is to maintain a safe distance between the two vehicles throughout the simulation. The main component of ACC is the controller that adjusts the ego car’s velocity to keep the relative distance above a desired safety threshold.
    \item \textit{Chasing Cars (CC).} Adapted from Hu et al.~\cite{hu2000towards}, the model of this system simulates five vehicles where only the lead car is directly controlled via throttle and brake inputs, while the others follow using a predefined algorithm. The system outputs the positions of all five cars. 
\end{itemize}

Our study subjects are specified in Simulink and are associated with a total of $15$ system requirements. These requirements define the expected behavior of each system model, including the input variables, their data types, and valid value ranges, all extracted from the corresponding specification documents. All our study subjects operate on numeric, time-continuous input signals and produce outputs of the same structure. Step~1 of DeCaF identified violations in $13$ out of the $15$ total requirements. Since our goal is to generate and characterize explanations for failure scenarios, we retain only the violated requirements for evaluation to answer our research questions. In the remainder of this section, we detail the configuration settings used for each step of the approach described in Section \ref{sec:approach}.

For Step~1, we adopt the parameterization used in prior work\cite{ernst2022arch,song2022cyber} to generate and simulate the training data. We conduct $30$ executions per system-requirement for ACC, and $50$ for AT and CC. Each execution is capped at $300$ optimization iterations. Test inputs in the training data are generated using control-point encoding (as detailed in Section~\ref{sec:approach}), evaluated against the requirement using the robustness metric~\cite{abbas2014robustness}, and iteratively evolved through search-based optimization. We used Simulated Annealing with Monte Carlo sampling to explore the input space and guide the search toward failure-inducing inputs. We configure the techniques used in Step~2 of DeCaF as follows. For M5 method, we adopt the default implementation from the \verb|m5py| Python library, using variance reduction as the split criterion to optimize the fit of regression models at the leaves. We limit the minimum number of instances per leaf to $2$ to prevent overfitting and apply pruning to simplify the tree. For Random Forests, we use an ensemble of $100$ unpruned decision trees, each trained on a bootstrapped sample of the training data with a random subset of input features considered at each split using the square root of the total number of features, following common practice. We evaluate the ensemble using mean squared error (MSE). For Step~3, we configure the counterfactual generation parameters as follows. We set the mutation and crossover probabilities to $0.1$ and $0.5$, used a population size of $50$ over $50$ generations, and returned up to $7$ counterfactuals per instance. Candidate selection followed a multi-objective loss, with a diversity weight of $0.1$, and we fixed the random seed to $17$ for reproducibility, following prior guidance where applicable~\cite{mothilal2020explaining}. The algorithm is evaluated using the three techniques described in Section~\ref{sec:approach}. Pre-trained causal models from Step~2 of DeCaF are integrated into the Diverse Counterfactual Explanations (DiCE) framework~\cite{mothilal2020explaining} using the \texttt{sklearn} backend. RS is enabled with the method ``random'', the GA with the method ``genetic'', and the KD strategy with the method ``kdtree''. For Step~4 of DeCaF, we apply the same M5 model tree settings used in Step~2. We execute DeCaF across all case-study systems using the configured techniques and evaluate it by conducting validation experiments to answer the three research questions. The next subsection presents these experiments and analyzes the results, ensuring fair comparisons across models and counterfactual generation strategies. 

\subsection{RQ1 - Effectiveness of DeCaF's Counterfactuals}

This question evaluates DeCaF’s effectiveness in generating counterfactual-guided explanations for CPS debugging, specifically its ability to generate valid counterfactuals that successfully restore requirement satisfaction for the system under test and its failing inputs. We start by empirically assessing the performance of DeCaF’s causal model learner, comparing four machine learning techniques, namely, M5, RIPPER, SVM, and Random Forest (RF), across all case-study systems. Based on this assessment, we select the best-performing techniques to evaluate the success rate of DeCaF quantifying the ability to generate valid counterfactual-guided explanations that flip the system output's verdict from a requirement violation to satisfaction. In the following, we analyze the results of both steps and conclude the main insights from our findings.

\subsubsection{Performance of Causal Model Learner} 

\begin{figure}[t]
    \centering    \includegraphics[width=\columnwidth]{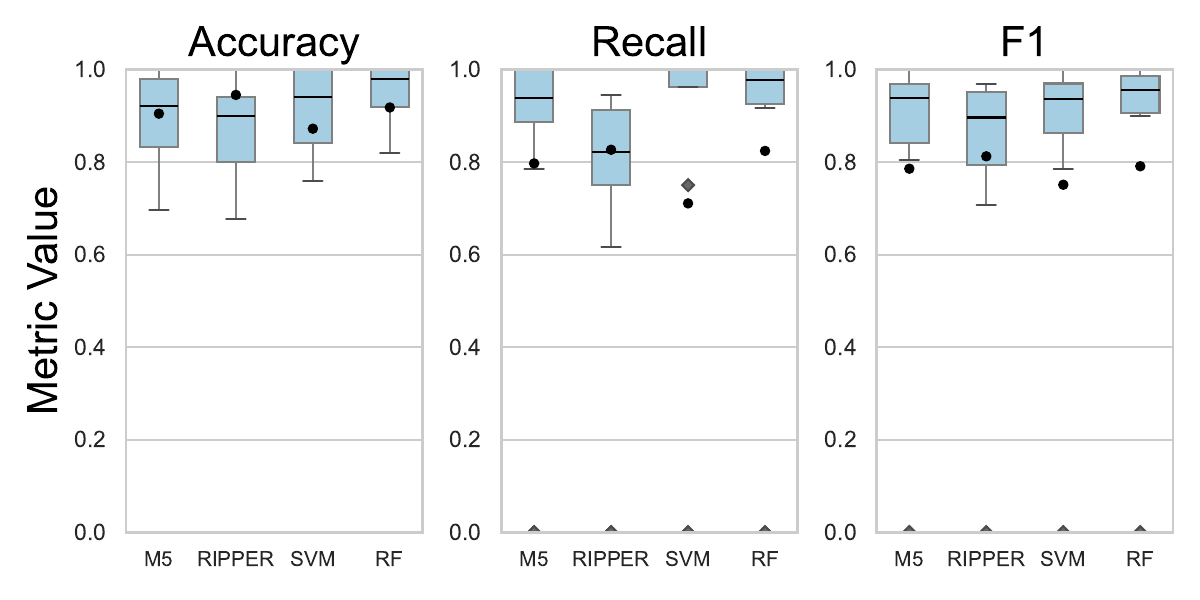}
    \caption{Distribution of evaluation metrics of the ML techniques across our systems}
    \label{fig:MLAccuracy}
\end{figure}

For each system–requirement, we use DeCaF's Step~1 (Section \ref{sec:approach}) to generate a set of test inputs. Each test input is then labeled as \texttt{pass} or \texttt{fail} based on whether it satisfies or violates the corresponding requirement. Table~\ref{tab:num_cf} presents the total number of test inputs (denoted as \emph{TS Size}) and the number of failing test inputs (denoted as \emph{\#Fail}) for each system. Using the four ML models, we train a causal model of system behavior on the labeled test input dataset, as input to DeCaF's Step~2. For each inferred causal model, we compute and analyze the \emph{Accuracy}, \emph{Recall}, and \emph{F1 Score} metrics. Figure~\ref{fig:MLAccuracy} shows the box plots of the distribution of metric values aggregated over $13$ system requirements in total. The black circular marker on each box denotes the average value for that model's metric combination.

Results show that among all classifiers, M5 consistently achieves the highest performance across all three metrics, closely followed by RF. Specifically, both models achieve high average accuracy ($0.90$ for M5 and $0.92$ for RF) and F1 scores ($0.79$ for M5 and $0.79$ for RF). This indicates M5 and RF's strong performance across our case-study systems. RF shows a tighter interquartile range and lower bottom whisker, indicating greater robustness and fewer performance outliers across system requirements. In contrast, SVM performs the worst overall ($0.87$ accuracy, $0.71$ recall, $0.75$ F1), while RIPPER exhibits a wide spread across the metrics, which suggest inconsistent classification of failure cases. Based on these results, we select M5 and RF as representative ML models to address the research questions of this paper. 

\subsubsection{Success Rate of DeCaF's Counterfactuals} To evaluate the success of DeCaF at generating valid counterfactual-guided explanations, we considered six configurations of DeCaF by combining three counterfactual generation strategies, namely RS, GA and KD with two ML models, namely M5 and RF that we selected as representative ML techniques. For each system requirement, the causal models were trained on the labeled test input dataset before passing them to Step~3 of DeCaF. 

\begin{table}[t]
\centering
\caption{Number of generated counterfactuals (\textit{\#CF}) for each system and DeCaF configuration. \textit{TS Size} denotes the total number of test inputs generated by Step~1, \textit{\#Fail} denotes the number of failing test inputs that violate the system requirement. Percentages denote \textit{\#CF} as a fraction of \textit{\#Fail}.}
\label{tab:num_cf}
\scriptsize
\setlength{\tabcolsep}{2pt}
\renewcommand{\arraystretch}{0.8}
\begin{tabular}{p{0.25cm} c c|cc|cc|cc}
\toprule
\textbf{Sys.} & \textbf{TS} & \textbf{Fail}
& \multicolumn{2}{c|}{\textbf{GA}} 
& \multicolumn{2}{c|}{\textbf{KD}} 
& \multicolumn{2}{c}{\textbf{RS}} \\
\cmidrule(lr){4-5}\cmidrule(lr){6-7}\cmidrule(lr){8-9}
& & & \textbf{M5} & \textbf{RF} & \textbf{M5} & \textbf{RF} & \textbf{M5} & \textbf{RF} \\
\midrule
AT  & 300 & 169 & 84 (50\%)  & 169 (100\%) & 166 (98\%) & 169 (100\%) & 166 (98\%) & 167 (99\%) \\
CC  & 250 & 177 & 0 (0\%)    & 177 (100\%) & 126 (71\%) & 177 (100\%) & 126 (71\%) & 174 (98\%) \\
ACC & 120 & 56  & 55 (98\%)  & 56 (100\%)  & 55 (98\%)  & 56 (100\%)  & 55 (98\%)  & 55 (98\%) \\
\midrule
\textbf{Tot.} & \textbf{670} & \textbf{402}
& \textbf{139 (35\%)} & \textbf{402 (100\%)} 
& \textbf{347 (86\%)} & \textbf{402 (100\%)} 
& \textbf{347 (86\%)} & \textbf{396 (99\%)} \\
\bottomrule
\end{tabular}
\end{table}

As shown in Table~\ref{tab:num_cf}, the total number of generated test inputs (see Total \emph{TS Size}) is $670$, of which $402$ are failing tests that violate system requirements (see Total \emph{\#Fail}). DeCaF’s Step~3 targets these failing inputs to generate counterfactual-guided explanations. To evaluate DeCaF’s applicability, we record the number of generated counterfactuals (\emph{\#CF}) for each configuration and report it as a fraction of the total failing inputs. The table summarizes these results across all systems and DeCaF configurations, combining GA, KD, and RS with either the M5 or RF model.

Across systems, RF using both GA and KD achieved $100\%$ applicability, generating at least one counterfactual for every failing input, followed by RF with RS that scored $99\%$. In contrast, M5 showed low applicability when paired with GA and scored the lowest applicability with M5, failing to generate any counterfactuals for the CC system and achieving only $35\%$ on average across all systems. This low performance is attributed to the sensitivity of M5 model trees under evolutionary optimization. Specifically, M5 models define piecewise functions with sharp decision boundaries that respond non-smoothly to small perturbations in the input space. Genetic algorithms, which rely on stochastic population-based operations such as mutation and crossover, can struggle to make consistent progress when confronted with such discontinuities, which prevents the algorithm from converging toward valid counterfactuals. Counterfactual generation techniques using the RF model were consistently more robust and achieving near-complete coverage across all the failing test inputs. These results confirm that both the counterfactual generation strategy and the underlying ML model significantly influence DeCaF’s ability to generate valid counterfactual-guided explanations.

\begin{table}[t]
\centering
\caption{Success rates and relative success rates of DeCaF's combinations of counterfactual generation strategies and ML models across our systems.}
\label{tab:res_success_rate}
\scriptsize
\setlength{\tabcolsep}{2.1pt}
\renewcommand{\arraystretch}{0.95}
\begin{tabular}{l|cccccc|cccccc}
\toprule
& \multicolumn{6}{c|}{\textbf{Succ. (\#Valid/\#Fail)}}
& \multicolumn{6}{c}{\textbf{Rel. Succ. (\#Valid/\#CF)}} \\
\cmidrule(lr){2-7}\cmidrule(lr){8-13}
\textbf{Sys.} & \multicolumn{2}{c}{\textbf{GA}} & \multicolumn{2}{c}{\textbf{KD}} & \multicolumn{2}{c|}{\textbf{RS}}
              & \multicolumn{2}{c}{\textbf{GA}} & \multicolumn{2}{c}{\textbf{KD}} & \multicolumn{2}{c}{\textbf{RS}} \\
\cmidrule(lr){2-3}\cmidrule(lr){4-5}\cmidrule(lr){6-7}
\cmidrule(lr){8-9}\cmidrule(lr){10-11}\cmidrule(lr){12-13}
& \textbf{M5} & \textbf{RF} & \textbf{M5} & \textbf{RF} & \textbf{M5} & \textbf{RF}
& \textbf{M5} & \textbf{RF} & \textbf{M5} & \textbf{RF} & \textbf{M5} & \textbf{RF} \\
\midrule
AT  & 0.46 & 0.77 & 0.93 & 0.51 & 0.79 & 0.67 & 0.92 & 0.77 & 0.95 & 0.51 & 0.81 & 0.68 \\
CC  & 0.00 & 0.73 & 0.59 & 0.65 & 0.42 & 0.26 & --   & 0.73 & 0.83 & 0.65 & 0.59 & 0.26 \\
ACC & 0.61 & 0.61 & 0.93 & 0.68 & 0.43 & 0.45 & 0.62 & 0.61 & 0.95 & 0.68 & 0.44 & 0.45 \\
\midrule
\textbf{Avg.} & \textbf{0.36} & \textbf{0.70} & \textbf{0.82} & \textbf{0.61} & \textbf{0.55} & \textbf{0.46}
              & \textbf{0.77} & \textbf{0.70} & \textbf{0.91} & \textbf{0.61} & \textbf{0.61} & \textbf{0.46} \\
\bottomrule
\end{tabular}

\vspace{0.3ex}
{\scriptsize \textit{Note:} ``--'' indicates configurations where DeCaF failed to generate any counterfactuals.}
\end{table}

Table~\ref{tab:res_success_rate} reports the success rates and relative success rates of DeCaF across various combinations of counterfactual generation strategies (GA, KD, RS) and ML models (M5, RF). The \emph{Success Rate} metric is defined as the proportion of failing test inputs for which at least one valid counterfactual was generated that satisfies the requirement \emph{(\#Valid CF / \#Fail)}. The \emph{Relative Success Rate} metric captures the success rate achieved by DeCaF when applicable, computed as the proportion of valid counterfactuals among all generated counterfactuals \emph{(\#Valid CF / \#CF)}. Each row corresponds to the rate on all aggregated system-requirements of a given system.  \emph{Average} indicates the average rate across all system requirements for each configuration. We consider RS as the baseline counterfactual generation method for DeCaF. Results show that RS performs the worst across both model types, with average success rates of $0.55$ for M5 and $0.46$ for RF, and relative success rates of $0.61$ and $0.46$, respectively. This indicates RS’s susceptibility to local optima and highlights the difficulty of navigating the high-dimensional input space without guided search and the need for better generation strategies to effectively generate valid counterfactuals.

Among all configurations, KD combined with M5 achieved the highest average success rate $0.82$ and relative success rate $0.91$, showing that M5 performs well when it is applicable or when inapplicability is minimal. This high effectiveness suggests that KD's local neighborhood exploration paired with the M5 model offers higher effectiveness in explaining the input space. RF which does not suffer from applicability issues performed reasonably well when paired with GA ($0.70$) and KD ($0.61$).

\begin{table*}[t]
\centering
\caption{Statistical comparison of DeCaF configurations. Left: (RS vs KD) and (RS vs GA). Right: impact of the counterfactual generation strategies and ML models (KD vs GA). $\hat{A}_{12}$ indicates effect size.}
\label{tab:compact_merged}
\scriptsize
\setlength{\tabcolsep}{2.0pt}
\renewcommand{\arraystretch}{0.92}
\begin{tabular}{llccl|llccl}
\toprule
\multicolumn{5}{c}{\textbf{RS vs KD and GA}} & \multicolumn{5}{c}{\textbf{KD vs GA}} \\
\cmidrule(lr){1-5}\cmidrule(lr){6-10}
\textbf{Sys.} & \textbf{Comp.} & \textbf{$p$} & $\boldsymbol{\hat{A}_{12}}$ & \textbf{ES}
& \textbf{Sys.} & \textbf{Comp.} & \textbf{$p$} & $\boldsymbol{\hat{A}_{12}}$ & \textbf{ES} \\
\midrule
\textsc{AT}  & RS vs GA           & 0.014                 & 0.538 & None
            & \textsc{AT}  & KD-M5 vs GA-M5         & 0.188                 & 0.485 & None \\
\textsc{CC}  & RS vs KD           & \textbf{$<10^{-15}$}  & 0.661 & Med.
            & \textsc{AT}  & KD-M5 vs KD-RF         & $<10^{-18}$           & 0.282 & Large \\
\textsc{CC}  & RS vs GA           & \textbf{$<10^{-12}$}  & 0.664 & Med.
            & \textsc{AT}  & KD-M5 vs GA-RF         & $2.1\times 10^{-6}$   & 0.412 & Small \\
\textsc{ACC} & RS vs KD           & \textbf{$<10^{-8}$}   & 0.683 & Med.
            & \textsc{CC}  & KD-M5 vs KD-RF         & 0.0015                & 0.361 & Med. \\
\textsc{ACC} & RS vs GA           & \textbf{$<10^{-8}$}   & 0.687 & Med.
            & \textsc{CC}  & KD-M5 vs GA-RF         & 0.025                 & 0.452 & None \\
\textsc{ALL} & (M5) RS vs KD      & \textbf{$<10^{-13}$}  & 0.617 & Small
            & \textsc{ACC} & KD-M5 vs KD-RF         & $<10^{-8}$            & 0.276 & Large \\
\textsc{ALL} & (RF) RS vs GA      & \textbf{$<10^{-14}$}  & 0.632 & Small
            & \textsc{ACC} & KD-M5 vs GA-RF         & $1.1\times 10^{-5}$   & 0.331 & Med. \\
           & Comb. RS vs KD       & \textbf{$<10^{-12}$}  & 0.589 & Small
            &            &                         &                       &       &      \\
\bottomrule
\end{tabular}
\end{table*}

We conducted a statistical comparison based on the validity of generated counterfactuals across our three case-study systems (AT, CC, ACC). We use the Mann–Whitney U test~\cite{mann1947test} for statistical significance and report Vargha–Delaney’s $\hat{A}{12}$ as an effect size measure~\cite{vargha2000critique}, and p-values for each pairwise comparison of RS against GA and KD, both at the individual system level and aggregated across all systems. The statistical analysis shows that RS consistently underperforms both KD and GA across our case-study systems, model types, and the aggregated dataset. On \textsc{CC} and \textsc{ACC}, RS is significantly outperformed by KD and GA ($p < 0.0001$) with medium effect sizes ($\hat{A}_{12}$ $>$ 0.66). On \textsc{AT}, the difference with GA is also significant ($p = 0.014$), though with a smaller effect ($\hat{A}_{12} = 0.538$). When combining all causal models (M5 and RF) across all systems, RS remains significantly worse than both KD and GA (e.g., $p < 0.0001$ and $\hat{A}_{12} \approx 0.63$ for GA using RF). Although effect sizes are mostly small to medium, their statistical significance and consistency confirm that RS underperforms. As an uninformed baseline, RS is meaningfully outpaced by guided methods like KD and GA. Next, we assess the impact of the counterfactual generation strategy (KD vs GA) and causal model choice (M5 vs RF) on DeCaF’s effectiveness. Comparisons between KD and GA under the same model (e.g., \textsc{AT}) yield non-significant p-values (e.g., $p = 0.18$), suggesting no clear winner when the model is fixed. In contrast, model choice has a stronger influence. For most systems, KD performs significantly better with M5 than RF (e.g., $p < 10^{-18}$, $\hat{A}_{12} = 0.28$, large), indicating that M5 enhances KD’s effectiveness. GA, however, shows no such sensitivity to model choice, with high p-values and $\hat{A}_{12}$ values near $0.5$, implying consistent behavior across models. Finally, comparing KD-M5 with GA-RF—each method’s best-performing configuration—confirms KD-M5's superiority, with significant results across all systems and $\hat{A}_{12}$ consistently below $0.5$ (e.g., $0.33$ on \textsc{ACC}, medium effect).


\subsection{RQ2 - Goodness of DeCaF's Counterfactual-Guided Explanations} 

\begin{table}[t]
\centering
\caption{Goodness of DeCaF configurations: (necessity, sufficiency) scores across systems.}
\label{tab:res_goodness}
\scriptsize
\setlength{\tabcolsep}{2.2pt}
\renewcommand{\arraystretch}{0.95}
\begin{tabular}{l|cccccc}
\toprule
\textbf{Sys.} & \multicolumn{2}{c}{\textbf{GA}} & \multicolumn{2}{c}{\textbf{KD}} & \multicolumn{2}{c}{\textbf{RS}} \\
\cmidrule(lr){2-3}\cmidrule(lr){4-5}\cmidrule(lr){6-7}
& \textbf{M5} & \textbf{RF} & \textbf{M5} & \textbf{RF} & \textbf{M5} & \textbf{RF} \\
\midrule
AT  & (.42, .98) & (.62, 1.00) & (.53, --)  & (.71, --)  & (.77, 1.00) & (.88, 1.00) \\
CC  & (--, --)   & (.54, .93)  & (.13, .88) & (.16, .92) & (.81, .96)  & (.78, .96)  \\
ACC & (1.00, .93) & (1.00, .78) & (1.00, --) & (1.00, --) & (1.00, 1.00) & (1.00, 1.00) \\
\midrule
\textbf{Avg.} 
    & \textbf{(.71, .96)} & \textbf{(.72, .90)}
    & \textbf{(.55, .88)} & \textbf{(.62, .92)}
    & \textbf{(.86, .99)} & \textbf{(.89, .99)} \\
\bottomrule
\end{tabular}

\vspace{0.3ex}
{\scriptsize \textit{Note:} ``--'' indicates configurations where (i) DeCaF failed to generate any counterfactuals for necessity and (ii) all control points of counterfactuals are modified for sufficiency.}
\end{table}

To answer RQ2, we evaluate the \emph{goodness} of DeCaF’s counterfactual explanations by analyzing how well a counterfactual captures the minimal and causally sufficient conditions for changing a system’s behavior from failure to success. We assess the \emph{goodness} using two complementary properties: \emph{necessity} and \emph{sufficiency}. These properties have been used to evaluate counterfactual explanations~\cite{kommiya2021towards}. We adapt these notions to control-point-level modifications in CPS inputs as follows. The \emph{necessity} score quantifies how critical the system inputs changes imposed by the generated counterfactuals are, i.e., if removing any of the system input-related change will cause the system to revert to failure. We consider the valid counterfactuals identified in RQ1. For each valid counterfactual, we iteratively revert the modified control-point values corresponding to each system input variable —one at a time— back to their original (failing) values, and re-simulate the system. If the requirement is violated again, the input variable changes imposed by the generated counterfactual are deemed necessary. The necessity score is computed as the proportion of valid counterfactuals in which the changes to the system inputs are deemed necessary. A higher score indicates that the changes introduced by DeCaF were more essential for resolving the failure, with fewer unnecessary or redundant modifications. The \emph{sufficiency} score captures whether the modified inputs alone are adequate to repair the failure. To compute it, for each valid counterfactual, we keep the values of the individually modified control points and generate $50$ random values for the remaining input control-points (within the corresponding input range). Each test of the set of 50 variants is used to simulate the system and observe the test outcome (pass/fail). The sufficiency score is defined as the fraction of these variants that still satisfy the requirement. A higher score indicates that the proposed counterfactual is more sufficient in inducing the desired correct system behavior.

Table~\ref{tab:res_goodness} shows the goodness scores of DeCaF's counterfactuals. Among all configurations, RS consistently achieve the highest average scores, with RS–M5 and RS–RF reaching $(0.86, 0.99)$ and $(0.89, 0.99)$, respectively. These results are expected given RS's tendency to converge to local optima, based on the insights drawn in RQ1 (see Table~\ref{tab:res_success_rate}), where RS struggles with low success rates. RS produces highly conservative counterfactuals, typically involving very minimal perturbations to a single control point to resolve a failure. This narrow search behavior lead to high necessity and sufficiency scores. However, this comes at the cost of effectiveness and applicability since the generated counterfactuals are not as generalizable across the wider input space, which reduces their usefulness. In contrast, both GA and KD generate counterfactuals achieve optimal goodness scores. Although their average goodness scores are moderately lower than those of RS, they were able to generate more impactful interventions and significant gains in success rates. For example, KD–M5, which achieved the highest overall success rate, has a goodness score of $(0.55, 0.88)$. GA–RF offers a well-balanced configuration with a goodness score of $(0.72, 0.90)$, reflecting a good trade-off between goodness and effectiveness, which is important for optimal applicability in the context of CPS debugging, where identifying necessary and sufficient failure counterfactuals effectively can help isolate the root cause of a failure.

Along with each set of counterfactuals, DeCaF generates corresponding natural language explanations that describe, in plain terms, the input changes responsible for restoring the system's correct behavior. The explanations further specify the exact time intervals encoded by the modified control points, and the associated value changes. For example, a generated explanation may state: \emph{``Input throttle at time [50–100s] changed from 0.2 to 0.8.''} We manually reviewed the full set of generated explanations and found that all were meaningful, non-vacuous, and aligned with their corresponding formal counterfactuals, confirming their interpretability and correctness.

\subsection{RQ3 - Complexity vs. Generalizability of DeCaF's Success Assertions}

As described in Section~\ref{sec:approach}, the success assertions inferred in Step~4 are derived from the set of valid counterfactuals (the maximum count is $7$) generated for the failing test input. We evaluate the complexity versus generalizability trade-off of DeCaF's success assertions, which is a common dilemma in the assertion-based inference techniques~\cite{miningassumption}, using the following evaluation steps. 

\begin{table}[t]
\centering
\caption{Safety measure, average number of predicates, and generalization score of success assertions inferred by DeCaF across configurations.}
\label{tab:res_safety}
\tiny
\setlength{\tabcolsep}{2.2pt}
\renewcommand{\arraystretch}{0.88}
\begin{tabular}{l|cccccc|cccccc}
\toprule
& \multicolumn{6}{c|}{\textbf{Safety}} 
& \multicolumn{6}{c}{\textbf{Avg Pred./G-score}} \\
\cmidrule(lr){2-7}\cmidrule(lr){8-13}
\textbf{Sys.} 
& \multicolumn{2}{c}{\textbf{GA}} 
& \multicolumn{2}{c}{\textbf{KD}} 
& \multicolumn{2}{c|}{\textbf{RS}}
& \multicolumn{2}{c}{\textbf{GA}} 
& \multicolumn{2}{c}{\textbf{KD}} 
& \multicolumn{2}{c}{\textbf{RS}}\\
\cmidrule(lr){2-3}\cmidrule(lr){4-5}\cmidrule(lr){6-7}
\cmidrule(lr){8-9}\cmidrule(lr){10-11}\cmidrule(lr){12-13}
& \textbf{M5} & \textbf{RF} & \textbf{M5} & \textbf{RF} & \textbf{M5} & \textbf{RF}
& \textbf{M5} & \textbf{RF} & \textbf{M5} & \textbf{RF} & \textbf{M5} & \textbf{RF} \\
\midrule
AT  & 0.75 & 0.65 & 0.85 & 0.87 & 0.99 & 0.96 & 1.75/1.00 & 2.35/0.99 & 1.84/1.00 & 2.54/1.00 & 2.13/1.00 & 2.44/1.00 \\
CC  & --   & 0.77 & 0.72 & 0.73 & 0.82 & 0.80 & --/--      & 2.39/0.95 & 1.64/1.00 & 2.58/1.00 & 1.32/1.00 & 2.56/0.96 \\
ACC & 0.71 & 0.66 & 0.57 & 0.56 & 0.85 & 0.92 & 1.71/1.00 & 2.31/0.79 & 1.50/1.00 & 2.62/0.89 & 1.40/0.96 & 1.95/0.96 \\
\midrule
\textbf{Avg.}
& \textbf{0.73} & \textbf{0.69}
& \textbf{0.71} & \textbf{0.72}
& \textbf{0.89} & \textbf{0.89}
& \textbf{1.73/1.00} & \textbf{2.35/0.91}
& \textbf{1.66/1.00} & \textbf{2.58/0.96}
& \textbf{1.62/0.99} & \textbf{2.32/0.97} \\
\bottomrule
\end{tabular}

\vspace{0.3ex}
{\tiny \textit{Note:} ``--'' indicates configurations where DeCaF failed to generate any counterfactuals or assertions.}
\end{table}

We assess the syntactic complexity of the inferred success assertions using the number of predicates they contain. A predicate is a Boolean condition involving a variable and a constant, typically expressed as $c \sim r, where \sim \in \{ <, \leq, >,\geq, =, \neq\}$. For instance, the assertion $(c_{th,1} \leq 20) \wedge (c_{b,1} \leq 5)$ contains two predicates: $(c_{th,1} \leq 20)$ and $(c_{b,1} \leq 5)$. More predicates indicate a more complex assertion. To assess generalizability, we define the \textsc{G\_score} metric as the proportion of valid counterfactuals covered by the inferred assertion. A highly generalizable assertion accurately captures all relevant counterfactuals associated with a given failing test input. An important trade-off exists between complexity and generalizability. Simpler assertions with fewer predicates tend to be easier to interpret, understand, and act upon during the debugging process, and they often generalize better across the input space. However, overly simple assertions may risk overfitting to outliers that violate the requirement. Conversely, more complex assertions can provide greater precision and insight into the specific input changes required to restore correctness, but may suffer from reduced interpretability and underfitting to the training data.

We evaluate which configuration can balance between these competing factors the best by generating assertions that maintain a reasonable degree of complexity and informativeness while ensuring sufficient generalizability. Table~\ref{tab:res_safety} reports the average number of predicates in the success assertions inferred from valid counterfactuals for our case-study systems across all DeCaF's configurations. Configurations using M5 consistently produce simpler (less complex) assertions than those using RF, averaging between $1.6$ and $1.7$ predicates versus $2.3$ to $2.6$ predicates for RF. The results of \textsc{G\_score} for the success assertions inferred from valid counterfactuals for our systems across all DeCaF's configurations show that all configurations with M5 achieve the highest \textsc{G\_score} with $0.996$ on average, while configurations with RF follow with $0.946$. This suggests that M5-based configurations generate success assertions that are syntactically less complex and more generalizable. Notably, KD with M5, which also achieves the highest success rates, yields one of the least complex assertions, which contributes to broader coverage of the input domain and may lead to lower necessity scores, as confirmed in RQ2. In contrast, GA with RF generates some of the most complex assertions, which more tightly constrain the input domain. This added complexity appears to increase sufficiency scores addressed in RQ2, without significantly compromising success rates. In summary, GA with RF achieves a more balanced trade-off across the metrics defined in the evaluation.

We further evaluate the capability of the inferred assertions to accurately characterize the region of correct behavior and avoid covering failing behaviors using the safety metric, which measures the proportion of new test inputs sampled within the assertion’s bounds that satisfy the requirement. Specifically, for each failing test input with a valid counterfactual generated by DeCaF using the six configurations, we generate $50$ new inputs by perturbing only the changed control points within the bounds of the assertion while holding other inputs fixed. High safety scores indicate that the assertion accurately characterizes the region of correct behavior and avoids covering failing behaviors, thus mitigating overfitting. Lower safety scores do not necessarily invalidate the assertion but suggest limited generalization, prompting engineers to inspect supporting counterfactuals for further insights.
As shown in Table~\ref{tab:res_safety}, all DeCaF configurations demonstrate high safety overall. Notably, RS configurations achieve the highest safety scores, which we attribute to their lower number of valid counterfactuals and tendency to converge on conservative (local) regions of the input space. However, this conservative behavior limits effectiveness and goodness, as evidenced in RQ1 and RQ2. In contrast, GA and KD configurations sacrifice roughly $15$–$20$\% safety compared to RS for significantly higher coverage, success and goodness rates, yielding more robust assertions.

\textit{Threats to Validity.} DeCaF uses stochastic procedures so outcomes from a single run can vary due to randomness; we mitigate this by repeating each system requirement experiment multiple times and reporting aggregate results. Performance can also depend on hyperparameter choices for the search procedures and learning models; we relied on default settings in the used libraries or values adopted from established prior work, which provides a consistent baseline but may not be optimal for every system. Regarding generalizability, our evaluation uses three widely-known Simulink CPS benchmarks with diverse STL requirements that are commonly used in CPS. We recognize the concern of using established ML learners. Our primary contribution is the framework itself; in the evaluation, we instantiate DeCaF with well-understood methods to demonstrate its viability. Finally, inputs are encoded as piecewise constant~\cite{ernst2021arch}, which is standard in CPS testing but may require adaptation for different domains.

%% file: Sections/related.tex
Simulation-based testing~\cite{10.1145/3711906,khatiri2023simulation,arrieta2019pareto} is widely adopted in the model-based design workflows of CPS through simulation tools such as Simulink~\cite{chaturvedi2017modeling}, which support automatic code generation. Prior work on automated failure diagnostics~\cite{sampath2002failure} in these environments has largely focused on model- and signal-level fault analysis, including trace monitoring and diagnostics~\cite{ferrere2015trace,bartocci2018localizing}. For example, Bartocci et al.~\cite{bartocci2021cpsdebug} analyze failing traces to identify anomalous Simulink components and fault-propagation paths, while Deng et al.~\cite{deng2023causal} employ causal temporal logic to explain faults in Simulink/Simscape models. However, these techniques do not provide input-level explanations that reveal the specific conditions under which a failure is triggered or could be avoided. The closest related effort is CaSTL~\cite{CaSTL}, which integrates causal theory with signal temporal logic to generate fault explanations in Simulink-modeled CPS. CaSTL synthesizes formulas that explain why a failure occurred using both the system model and fault observation traces. While this represents a significant advancement in model-based causal explanation, it does not generate input-level interventions that would restore correctness. In contrast, our work shifts the focus to counterfactual input-level changes, identifying precisely what modifications to test inputs would have prevented the failure.

Delta debugging~\cite{zeller2002isolating} isolates failure inducing circumstances by repeatedly minimizing a failing input while preserving the failure. Statistical debugging and spectrum based fault localization similarly narrow the search by ranking suspicious elements. For Simulink, Liu et al.~\cite{liu2016simulink} connect failing tests to model outports and blocks using execution slicing and SBFL, and use supervised learning to cluster related faults. These techniques help localize where failures manifest, but they typically do not identify triggering time intervals, provide causal explanations, or support intervention based reasoning for recovery. In contrast, DeCaF generates simulator validated repaired inputs and summarizes them as an interpretable success region over the system inputs.

Assertions inference has also been investigated to characterize correctness conditions. Daikon~\cite{ernst2007daikon} dynamically infers program invariants, while tools such as Alhazen~\cite{kapugama2022human, soremekun2020inputs} mine grammar-based assertions for predicting failures. Other work derives environment assumptions for CPS~\cite{kampmann2020does, gopinath2020abstracting, gaaloul2021combining, miningassumption, enrich}. The closest approaches~\cite{gaaloul2021combining, miningassumption} to DeCaF’s step~4 characterize pass/fail conditions directly on time-indexed CPS input signals. Whereas those works evolve predicates via genetic programming, we infer them with M5 model trees, for best trade-offs between complexity, generalizability, and safety. Additionally, these works lack causal reasoning as they don't support counterfactual interventions to restore correctness.

%% file: Sections/conclusion.tex
This paper introduced DeCaF, a framework for debugging CPS failures through input level counterfactual explanations. Rather than only localizing faults to internal components, DeCaF identifies minimal and precisely timed input signal changes that are necessary and sufficient to prevent failures, generating counterfactuals via random search, genetic algorithms, and KD tree nearest neighbors. Across three automotive case studies, DeCaF consistently produced high quality counterfactuals and inferred concise, generalizable, human understandable success assertions that characterize recovery conditions.